\newif\ifarxiv
\arxivtrue 
\ifarxiv
	\documentclass[12pt]{article}
	\usepackage[margin=1in]{geometry}
	\usepackage[slantedGreek]{mathpazo}
	\usepackage[pdftex]{graphicx}
	\usepackage{amsfonts}
	\usepackage{amssymb}
	\usepackage{amsthm}
	\usepackage{graphicx,float}
	\usepackage{amsmath}
	\usepackage{hyperref}
	\usepackage{natbib}
	\usepackage{breakcites}
\else
	\documentclass[authoryear,12pt]{elsarticle}
	\usepackage{amsfonts}
	\usepackage{amssymb}
	\usepackage{amsthm}
	\usepackage{graphicx,float}
	\usepackage{amsmath}
	\usepackage{hyperref}
	\usepackage[margin=1in]{geometry}
\fi

\usepackage{color,marginnote}
\usepackage{array}
\usepackage{amsmath}

\newcommand{\defeq}{\mathrel{\mathop:}=}
\newcommand{\st}{\mbox{subject to}\quad}
\newcommand{\E}[1]{\operatorname{E}\left(#1\right)}
\newcommand{\Var}[1]{\operatorname{Var}\left(#1\right)}

\DeclareMathOperator*{\minf}{minimize \quad}
\DeclareMathOperator*{\maxf}{maximize \quad}

\DeclareMathOperator*{\argmax}{arg\,max}

\newcommand{\mindisp}[3]{\begin{align*}\minf_{#1} & #2\\\st  & #3\end{align*}}

\ifarxiv
\title{Strategic Bayesian Asset Allocation}
\author{
	Vadim Sokolov\thanks{Vadim Sokolov is Assistant Professor in Operations Research at George Mason University. email:vsokolov@gmu.edu}~\ and Michael Polson\thanks{Michael Polson as an Analyst at Bates White. email:michael.alan.polson@gmail.com }
}
\date{First Draft: November 2018\\This Draft: November 2019}
\else
\fi

\graphicspath{{./fig/}}

\begin{document}

\ifarxiv
\maketitle
\begin{abstract}
\noindent
Strategic asset allocation requires an investor to select stocks  from a given basket of assets. The perspective of our investor is to maximize risk-adjusted alpha returns relative to a benchmark index. Historical returns are used to provide inputs into an optimization algorithm. Our approach uses Bayesian regularization to not only provide stock selection but also optimal sequential portfolio weights. By incorporating investor preferences with a number of different regularization penalties we extend the approaches of \cite{black1992global} and \cite{puelz2015sparse}. We tailor standard sparse MCMC algorithms to calculate portfolio weights and  perform selection. We illustrate our methodology on stock selection from the SP100 stock index and from the top fifty holdings of two hedge funds Renaissance Technologies and Viking Global. Finally, we conclude with directions for future research. 
\end{abstract}
 \fi


\section{Introduction}
Strategic asset allocation requires an investor to select stocks from a given basket of assets with a goal to outperform a benchmark index by a given margin of alpha. We propose to use  Bayesian regularization, based on predicted expected returns and volatility from historical returns, to simultaneously perform asset selection and optimal portfolio allocations. Our approach builds on the  asset  allocation model of \cite{black1992global} which has become the gold standard for practitioners. This approach incorporates investors' preferences on expected returns. \cite{puelz2015sparse} provide a similar analysis using shrinkage and selection to select stocks using regularized regression.


\tolerance=300
Our methodology builds on traditional stock selection  techniques in a number of ways. First, we recast portfolio selection as a  regularization problem with a goal  to optimize expected returns subject to a desired number of stocks in the portfolio. Second, we use sparsity inducing prior distributions such as spike-and-slab, horseshoe, and Lasso to perform selection. Thirdly, the  output from our algorithm naturally orders assets to include into the optimal portfolio in a one-by-one fashion so that performance can be dynamically tracked over time.
Our contributions include 
\begin{itemize}
	\item Duality between hierarchical linear models and constrained optimization problem. Hierarchical linear model allows for more flexible set of prior distributions and allows to use existing MCMC algorithms for finding optimal portfolios
	\item Duality between uncertainty in input data and allocation weights. This provides a new interpretation of sparsity inducing the Laplace prior which corresponds to Lasso estimator. 
	\item Extending the Black-Litterman framework to sparsity-inducing priors
	\item Our empirical results demonstrating effectiveness of several sparsity-inducing priors on portfolio selection and its performance. 
\end{itemize}

The traditional \cite{markowitz1952portfolio}  mean-variance approach  is optimal under the assumption that historical means and variances provide good estimates of future expected returns and volatilities. A number of practical considerations, such as preferences to include as few stocks as possible into portfolio, needs further research. Markowitz portfolio selection is also known to be sensitive to errors in estimation and can lead to extreme weights which hinders its wide spread adoption. Black-Litterman use shrinkage (a.k.a. ridge regression) to improve performance and to allow for investor preferences. \cite{polson2000bayesian} use sparsity inducing priors that impose investors' beliefs and show how to estimate covariance matrices for large-scale portfolio problems. 

\cite{black1992global} proposed using Gaussian regularization to perform the shrinkage and frame this approach in the context of combining quantitative and subjective beliefs into a predictive model of expected returns. We show how to  incorporate sparsity and selection. While the Markowitz approach relies on convex optimization, BL relies on Bayesian posterior mode updates of portfolio weights, and as we will show later leads to identical solutions under certain assumptions. \cite{stevens1998inverse} provided the connection between statistical and mean-variance based approaches.  \cite{bertsimas2012inverse} show how the original BL approach can be recast as an optimization problem. 

Our contribution is both methodological and empirical. On the methodological side, we develop a new  optimal asset  allocation algorithm using sparsity. On the empirical side, we compute asset allocation and selection for both static buy-and-hold and dynamic optimal re-balancing. We evaluate three sparsity inducing priors for portfolio allocation, namely Laplace, Horseshoe \citep{carvalho2010horseshoe} and spike-and-slab. The empirical performance of sparsity approaches is demonstrated by comparing performance of selected portfolios with SP500 returns.  The LARS algorithm \citep{efron2004least} is particularly effective in finding the optimal portfolio  posterior mode for the Laplace prior. MCMC works well for Horseshoe \cite{hahn2019efficient}, and Single Best Replacement (SBR) \cite{polson2017bayesian} algorithm for finding posterior mode for spike-and-slab.

Our approach naturally addresses the problem of portfolio selection and the issue of over fitting is alleviated by applying regularization techniques, see \citet{kandel1995bayesian} who address the problem of distribution shifts with Bayesain techniques. \citet{carrasco2011optimal} apply regularization technique for the degeneracy of the estimated covariance matrix of a large number of assets.

\subsection{Connections with Previous Work}
Mean-variance portfolio selection has a long-standing place in financial econometrics with ground-breaking work  of \citet{de1940problema} and \citet{markowitz1952portfolio}. Scalability of optimization algorithms and incorporation of transactions costs was addressed by \citet{perold1984large}. A number of authors have provided Bayesian solutions \citep{barberis2000investing}, using a the predictive model of stock returns together with regularization.

Our work builds on other Bayesian stock selection strategies such as those based on factor modeling, see~\citet{black1992global,carvalho2011dynamic,aguilar2000bayesian,puelz2015sparse,getmansky2015hedge}. Our main assumption is predictability~\citet{barberis2000investing,kandel1996predictability}  of stock returns, which is the main justification  that an investor should heavily invest into stocks due to the large equity premium. A related approach is to regularize with $\ell_1$ penalties (lasso) which has been well studied in the  portfolio optimization literature, see for example \citep{Fastrich2015constructing}. $\ell_1$-based approaches have their shortcomings including over-shrinkage and the inability to recover sparse signals for highly dependent data. 

To illustrate this issue, several authors have proposed non-convex approaches. \cite{gasso2009recovering} and \cite{Giuzio2018} use $\ell_q$ penalties to address the issue of highly dependent data and allocate portfolios during a crisis. Other non-convex penalties include smoothly clipped absolute deviations (SCAD) \citet{fan2001variable} and its linear approximation \citet{zhang2009some}. Bridge \citep{polson2014bayesian} or $\ell_q$ penalty \citet{frank1993statistical} is a generalization of more widely used $\ell_1$ (LASSO) and $\ell_2$ (Ridge) penalties. \citep{bourgeron2018robust} show that the mean-variance optimization approach will play an arbitrage between assets with similar return and volatility and show that regularization techniques are required to find a robust asset allocations. 

A Bayesian approach naturally allows an investor to incorporate uncertainty about mean-variance parameters. For example, \citet{carvalho2011dynamic} address the problem of change in covariance estimates by dynamically updating it as new observations arrive \citep{jacquier2012asset}. Robust minimax  optimization techniques were recently proposed to account for uncertainty in covariance matrix estimator and to solve for the worst-case scenario, see \citet{ismail2019robust}. Our work builds on \citet{puelz2015sparse} who use Bayesian techniques to design a mean-variance portfolio with a small number of assets and analyze the trade off between optimality and number of assets to be included.  \citet{kozak2017shrinking} design sparse factor models for analysis of large number of cross-sectional stock returns.   \citet{jacquier2012asset} proposed decision-theoretic framework for asset allocations that relies on Bayesian analysis, see \cite{avramov2010bayesian} for further discussion.


\section{Strategic Bayesian Asset Allocation}
Traditional mean-variance portfolio optimization assumes that returns at time $t$ of asset $i$, $r_{it} = p_{it}/p_{i(t-1)}$ of each of $n$ assets follow a distribution with mean $\mu \in R^n$ and covariance matrix $\Sigma \in R^{n\times n}$. Here $p_{it}$ is the price of asset $i$ at time $t$.  Returns of each asset are assumed to be a weak-stationary stochastic process. 

The investor's objective is to minimize the variance (risk) of the portfolio while having a guaranteed return $\rho$. A portfolio is defined by a vector of weights (allocations), denoted by  $w = (w_1,\ldots,w_p)$. Thus, the variance of the portfolio is given by $w^T\Sigma w$ and the expected return by $\mu^Tw$.  

Then,  the optimal portfolio is found by solving the following optimization problem
\begin{align}
\minf_{w} & w^T\Sigma w \label{eqn:markowitz}\\\nonumber
\st  & r^Tw = \rho,~~~1^Tw =  1, ~~~w_i\ge 0 
\end{align}
where the weight $w_i$ is the amount of asset $i$ held throughout the period. 
Positivity constraint $w_i \ge 0$ are added to guarantee that only long positions are to be included.

\subsection{Extending Black-Litterman}
The first widely used regularization approach was proposed by \cite{black1992global}. The BL model uses quadratic regularization, which can be interpreted as a mechanism to integrate quantitative and traditional portfolio building strategies. The BL model assumes a normal prior over investor's beliefs over future returns. The objective function then combines loss minimization with the regularization term that encodes investors' beliefs. In other words, the BL model combines quantitative and traditional management approaches and allows to update currently held beliefs using observed data (returns) to form new opinions. 

We begin by providing a framework to determine an optimal portfolio allocation from the perspective of an informed investor. Let $\pi$ denote an expected return vector for the $n$ assets. Following Black-Litterman, let $\pi^*$ denote beliefs about expected returns over and above those implied by $\pi$. Expected returns $\pi$, are typically determined from an equilibrium model, such as CAPM or FF3/5. 

Specifically, under CAPM model , we have 
\[
\pi = \beta(\mu_m - r_f) \; \; {\rm where} \; \; \beta = \mathrm{Cov}(r,r^Tw_m)/\sigma_m^2 .
\]
Here $w_m^Tr$ is the return on the (global) market, $r_f$ is a risk-free rate, $\sigma_m^2$ is the market variance, and $r$ is the vector of asset returns. We can then write $\pi = \sigma \Sigma w_m$, where $\Sigma = \mathrm{Cov}(r,r^T)$. Here, $\sigma = (\mu_m - r_f)/\sigma_m$ is the Sharpe ratio of the market index. From this, we can  calculate optimal weights given expected returns, namely $w_m = (\sigma \Sigma)^{-1}\pi$. 

This duality allows us to move between weights and expected returns. Informed investor beliefs are typically given by a constraint of the form $P\pi^* = q$, where $P$ is a vector and $P_i$ identifies if there is an opinion about  asset $i$, and $Q$ is the vector of opinions. As a simple example, suppose that $P = (-1,1,0,\dots,0)$ and $q = (0.5,0,\dots,0)$ express the belief that the first asset will outperform the second by 0.5\% over the period in question. Uncertainty in this belief can be expressed as $P\pi^* \sim N(q,\tau\Omega)$, where $\tau$ measures the strength of such a belief. BL show how Bayesian updating lets the investor to calculate the distribution over the new beliefs $p(\pi^*\mid \pi)$
\[
\pi\mid \pi^* \sim N(\pi^*,\Sigma),~~~P\pi^* \sim N(q,\tau\Omega)
\]
combined with Bayes rule,
\[
\pi^*\mid \pi \sim N\left(\mu, \left((\tau\Sigma)^{-1} + P^T\Omega^{-1}P\right)^{-1}\right)
\]
\[
\mu = \left((\tau\Sigma)^{-1} + P^T\Omega^{-1}P\right)^{-1}\left((\tau\Sigma)^{-1}\pi + P^T\Omega^{-1}Q\right)
\]
Regularization is now a central tool to allow investors to perform stock selection. This requires a selection of a norm on the portfolio weights. Several authors have applied $\ell_1$ regularization. \citet{demiguel2009generalized}  build on the work of \citet{jagannathan2003risk} and \citet{ledoit2004well} and propose a general mean-variance portfolio allocation framework in which norm of portfolio weights is constrained. They show duality of constraint-based approach and Bayesian approach in which investor assigns prior distribution for each of the weights. \citet{lobo2007portfolio} show that inclusion of transaction costs make $\ell_1$ regularized formulation to be non-convex and propose convex relaxations that can be efficiency solved.    \citet{brodie2009sparse} showed that regularization techniques do improve predictive power of statistical models for stock portfolios. \citet{fan2012vast} show that regularized mean-variance approach allows to achieve similar performance to the theoretically optimal portfolio while using covariance matrix estimated from a sample.

\subsection{Portfolio Weight Regularization}
In order to apply Bayesian inference algorithms, we re-cast the optimization problem as a hierarchical Bayesian linear model. To do this, let the return of the portfolio at time $t$ is given by the return vector $r_t = (r_{1t},\ldots,r_{pt})$. Then, the empirical estimate for the variance of the portfolio is given by 
\[
	w^T\Sigma w = \Var{w^Tr} = \E{(r^Tw-\mu^Tw)(r^Tw-\mu^Tw)^T} = \E{(r^Tw-\rho)(r^Tw-\rho)^T}
\]
The empirical risk is given by $w^T\hat{\Sigma} w = (1/T)\sum_{t=1}^T (\rho - r_t^Tw)^2$. 

We re-write risk minimization objective as a least-squares problem
\mindisp{w}{(1/T)\lVert\boldsymbol{\rho} - Rw\rVert_2^2 }{\mu^Tw = \rho\\ &1^Tw =  1,}
were $\boldsymbol{\rho} = (\rho,\ldots,\rho) \in R^T$. 

The return matrix $R$ is typically ill-conditioned. This can lead to unstable numerical solutions of the above problem. It is a usual problem when assets are highly correlated, then the columns of matrix $R$ become almost linearly dependent and the matrix becomes ill-conditioned. One approach to stabilize the solution and to find sparse portfolios is to add a regularization penalty to the objective function \citet{brodie2009sparse}. \cite{brodie2009sparse} analyzed the case where the penalty function is based on absolute value $\phi(w) = \sum_i|w_i|$ and showed that it leads to a stable solution. Our approach builds on the work of \cite{puelz2015sparse} who viewed the absolute value penalty as a way to incorporate investor's desire for a simple portfolio. They take a similar view as Black and Litterman and show that investor's preference to allocated her wealth among a small number of assets.

Sparse portfolios allow us to reduce transaction costs by eliminating certain stocks and to minimize the number of stocks that an investor need to follow and research. Another, interpretation of the penalized objective~\cite{puelz2015sparse} is that it allows to incorporate investor's preference with regards to number of stocks to be included into portfolio
\mindisp{w}{(1/T)\lVert\boldsymbol{\rho} - Rw\rVert_2^2 + \tau\phi(w)}{\mu^Tw = \rho\\ &1^Tw =  1 }
Noting that we do not include the positivity constraint into our regularized formulation and allow for short positions in the portfolio. A regularization penalty added to the objective function allows to stabilize portfolio~\cite{brodie2009sparse}. Thus the positive weight constraint can be excluded in a regularized formulation.  In order to satisfy the $1^Tw =  1$ constraint, we modify the problem by subtracting first column of the $R$ matrix from other columns and then  estimate $p-1$ linear coefficients of the modified problem $w_2,\ldots,w_p$ and finally calculate the remaining weight $w_1 = 1- \sum_{i=2}^p w_i$.

Denote the penalized empirical risk now by $g(w) = (1/T)\lVert\boldsymbol{\rho} - Rw\rVert_2^2 + \tau\phi(w)$. The corresponding Lagrangian dual function associated with the optimization problem with the $1^Tw =  1$ constraint excluded is given by
\[
	h(\lambda) = \inf_{w} \left[g(w) + \lambda(\rho - \mu^Tw ) \right].
\]
The dual function yields lower bounds on the optimal portfolio $w^*$. For any $\lambda$, we have
\[h(\lambda) \le g(w^*).\]
For a specific value of $\lambda^* \in \argmax_{\lambda} g(\lambda)$, we have $h(\lambda^*) = g(w^*)$.
Since $\lambda \rho $ term does not depend on $w$, we re-write the problem as 
\begin{equation}\label{eqn:lagrange}
	\minf_{w} (1/T)\lVert\boldsymbol{\rho} - Rw\rVert_2^2 + \tau\phi(w) - \lambda \mu^Tw.
\end{equation}
We cam select the value of the dual variables $\lambda$ and $\tau$ using cross-validation or naturally as stocks get added to the portfolio.

The penalty terms $- \lambda \mu^Tw$ and  $\tau\phi(w)$ in (\ref{eqn:lagrange}) can also be viewed as judgement of an investor that need to be incorporated into portfolio allocation decision making. To interpret those terms as prior judgement, we re-write the optimization problem as a Bayesian inference problem. 


\subsection{Duality between weight and moment regularization}
Bayesian techniques for portfolio selection can be divided into two categories, those that assume priors on the expected returns and variance-covariance matrix and those that place priors on the portfolio weights. Under certain conditions regularizing priors on the weights are equivalent to approaches that assume uncertainty in the observed data.  For example, \citet{jagannathan2003risk} show that adding non-negative weight constraint is equivalent to shrinking elements of the covariance matrix, which leads to reduced risk portfolio and more stable allocations. This follows directly from the KKT (Karush-Kuhn-Tucker) optimality condition for the portfolio optimization problem with an additional constraints $0 \ge w_i \le \bar w$, so that
\begin{align*}
	\sum_j \Sigma_{ij}w_j - \gamma_i + \sigma_i & = 0,~~~i=1,\ldots,N\\
	\gamma_i \ge 0 &,~~ \gamma_i = 0,\mbox{ if } w_i >0\\
	\delta_i \ge 0 &,~~ \delta_i = 0 \mbox{ if } w_i < \bar w
\end{align*}

Here $\gamma$'s and $\delta$'s are Lagrange multipliers. Thus, solving constrained problem is equivalent to solving the unconstrained problem with $\tilde{\Sigma} = \Sigma + \Delta  - A$, where $\Delta_{ij} = \sigma_i + \sigma_j$, and $A_{ij} = \gamma_i + \gamma_j$. Thus, the Lagrange multipliers $\gamma$ associated with the $w_i \ge 0$ constraint are effectively shrink the elements of the covariance matrix $\Sigma$. More specifically, $\Sigma_{ij}$ is reduced by $\gamma_i + \gamma_j$, and $\Sigma_{ii}$ is reduced by $2\gamma_i$. 

More generally, \cite{xu2009robust} and \citep{bertsimas2016best} provide duality between robust optimization which assumes uncertainty over return data and linear models with weight regularization. Specifically, one can show that a robust min-max formulation of a maximum likelihood estimation for a linear model
\[
\minf_{w \in \mathbb{R}^P} \left(\maxf_{\Delta R \in U}\lVert \rho - (R + \Delta R)w \rVert_2^2\right)
\]
with feature-wise uncertainty set given by $U = \{\Delta R\mid \lVert \Delta R_i \rVert_2 \le c, i=1,\dots,p\}$
is equivalent to an $\ell_1$ penalized log-likelihood maximization
\[
\minf_w \lVert \rho - R w \rVert_2^2 + \sum_{i=1}^{p}c_i |w_i|.
\]


We now turn to stock selection. 

\subsection{Stock Selection via Bayesian Inference}
The key insight is that the optimization problem defined by (\ref{eqn:lagrange}) is equivalent to finding a mode of a posterior distribution for a linear Gaussian model with exponential prior on the parameters and sparsity prior (regularization). 
\begin{equation}
	\boldsymbol{\rho}  = Rw  + \epsilon, \mbox{ where } w \sim p(w)  f(w),~~ \epsilon \sim N(0,\sigma_e^2I).
	\label{eq:bayes}
\end{equation}
Here the prior is a product of factors with distributional assumptions $p(w)  =  \exp(-\lambda \mu^Tw)$ and $f(w)\propto \prod_{i=1}^p\exp(\tau \phi(w_i))$. 

Since $-\log p( \rho\mid R,w) \propto \lVert\boldsymbol{\rho} - Rw\rVert_2^2 $, the mode of the log-posterior distribution over the coefficients of the above linear model is given by	
\[
\log p(w\mid R,\boldsymbol{\rho} )  = c+  \log p( \boldsymbol{\rho} \mid R,w) + \log p(w) + \log f(w).
\]
This is equal to the solution of the optimization problem given by (\ref{eqn:lagrange}). The exponential prior corresponds to the equality constraint $\mu^Tw = \rho$.

The exponential prior $\exp(-\lambda \mu^Tw)$ is conjugate and the posterior can be analytically calculated as follows
\begin{align*}
p(w\mid R,\boldsymbol{\rho} ,\lambda) & \propto \exp\left(-\dfrac{1}{2}\lVert\boldsymbol{\rho}  - Rw\lVert^2 - \lambda \mu^Tw\right)\\
	       & \propto \exp\left(-\dfrac{1}{2}w^TR^TRw + w^TR^T\boldsymbol{\rho} - \lambda \mu^Tw\right)\\
	       & \propto \exp\left(-\dfrac{1}{2}\left(w - (R^TR)^{-1}(R^T\boldsymbol{\rho}  - \lambda\mu)\right)^TR^TR \left(w - (R^TR)^{-1}(R^T\boldsymbol{\rho}  - \lambda\mu)\right)\right).
\end{align*}
By combining the likelihood and exponential prior $p$, we get the normal posterior with mean $(R^TR)^{-1}(R^T\boldsymbol{\rho}  - \lambda\mu)$ and covariance $(R^TR)^{-1}$. The resulting linear model is 
\begin{equation}
	\tilde{\boldsymbol{\rho}}  = Rw  + \epsilon, \mbox{ where } w \sim f(w),~~ \epsilon \sim N(0,\sigma_e^2I).
	\label{eq:bayes1}
\end{equation}
where $\tilde{\boldsymbol{\rho}}  = R(R^TR)^{-1}(R^T\rho- \lambda\mu) $. The corresponding optimization problem is then 

\begin{equation}\label{eq:lagr}
	\minf_{w} \lVert \tilde{\boldsymbol{\rho}} - Rw\rVert_2^2  + \tau\phi(w)
\end{equation}

Sparsity-inducing prior $f(w)$ and the corresponding penalty $\phi(w)$ leads to stable numerical solution robust to estimation errors in covariance and allow for sparse portfolios. 

Since matrix $R$ can be ill-conditioned in practice, we use a QR-decomposition to calculate $\tilde{\rho}$ which has better numerical properties compared to a LU decomposition \cite{golub1996matrix}. Further, as we have discussed earlier by adding regularization terms for the weight parameters is equivalent to assuming uncertainty in the input data and thus is robust to perturbations in the $R$ matrix.

\section{Sparsity-Inducing Portfolio Priors}
Bayesian formulation of portfolio selection problem allows to gain insight and to provide an alternative interpretation of the constraints and the corresponding penalty terms. Additionally, it allows to quantify uncertainty over the portfolio weights. Fully Bayesian inference that relies on MCMC algorithms allows to calculated credible intervals and thus to assess uncertainty. Efficient MCMC algorithms can be constricted by exploiting latent variable tricks. For example, the $\ell_1$ penalty $\phi(w) = \lVert w\rVert_1 = \sum_{i=1}^n|w_i|$ corresponds to Laplace prior distribution $f(w_i) =  (1/(2b))\exp(-|w_i|/b)$. Using a latent variable trick allows us to re-write this prior as a scale mixture of normals~\citep{andrews1974scale,west1987scale, carlin1991inference}. 

Specifically,  introduce latent variable $\gamma_i$ with an exponential distribution,
\begin{align*}
w_i \mid \sigma^2,\gamma \sim &N(0,\gamma\sigma^2)\\
\gamma  \mid \lambda \sim & Exp (\lambda^2/2)\\
\sigma^2 \sim & \pi(\sigma^2).
\end{align*}
There is an equivalence with the $\ell_1$ penalty obtained by integrating out $\gamma$
\[
p(w_i\mid \sigma^2,\lambda) =  \int_{0}^{\infty} \dfrac{1}{\sqrt{2\pi \gamma\sigma^2}}\exp\left(-\dfrac{w_i^2}{2\sigma^2\gamma}\right)\dfrac{\lambda^2}{2}\exp\left(-\dfrac{\lambda^2\gamma}{2}\right)d\gamma = \dfrac{\lambda}{2\sigma}\exp(-\lambda |w_i|/ \sigma ).
\]
In general, many widely used priors can be represented as variance-mean mixtures, using latent variable. The resulting model is linear with heteroscedastic errors~\citep{polson2013data}:
\[
p(w_i \mid \lambda) = \int_{0}^{\infty}\phi(w_i\mid \mu + \kappa\gamma_i^{-1},\lambda^2\gamma_i^{-1})p(\gamma_i)d\gamma_i,
\]
where $\phi(a\mid m,v)$ is the density function of normal variable with mean $m$ and variance $v$. 


\subsection{Elastic Net Priors}
The original BL \cite{black1992global} framework combines model-driven and traditional portfolio management into in a joint framework. Simply put, it allows us to combine predictions from quantitative models together with with investor beliefs.  

We modify this approach and introduce a sparse Black-Litterman selection. Our Bayesian methodology updates opinions of an investor with model-based predictions to form a new set of predictions while maintaining sparse portfolio. 
\[
p(w_i \mid \lambda_1 , \lambda_2 , \sigma^2 ) =  \dfrac{\lambda_1}{2\sigma}\exp(-\lambda_1 |w_i|/ \sigma ) \dfrac{\sqrt{\lambda_2} }{\sqrt{2 \pi} \sigma}\exp(-\frac{\lambda_2}{\sigma^2} w_i \Omega^{-1} w_i).
\]
Here we have two regularization parameters $ \lambda_1 $ and $ \lambda_2 $ to choose.  The parameter $\lambda_2$ is set proportional to the strength of the investor's beliefs  in their side-information and not bu the cross-validation. A Gibbs sampler can be constructed by representing the Laplace prior as a scale mixture of normal \cite{carlin1991inference}.  For an alternative empirical Bayes approach to select values of the hyperparameters for cases when design matrix is highly collinear see \cite{liu2018bayesian}.

The posterior MAP estimate can be computed as a fast convex optimization problem by combining the two  normal terms and then using the LARS algorithm. This provide a scalable version of the Black-Litterman that allow us to perform selection. Similar approach was considered by \cite{ho2015weighted} who shoed that elastic net regularization allows to improve out-of-sample performance of the mean-variance portfolio. We can still use LARS and thus see how stocks are sequentially added (without finding an optimal $\lambda_1$ by cross-validation).

\subsection{Horseshoe Priors}
Horseshoe priors belongs to a class of global-local class of priors and is defined by global parameter $\tau$ that does not depend on index $i$ and local parameter $\lambda_i$ which is different for each parameter $w_i$. The prior is defined by 
\[
w_i \mid \lambda_i,\tau \sim N(0,\tau^2\lambda_i^2).
\]
The global hyper-parameter $\tau$ shrinks all parameters towards zero. The prior for the local parameter $\lambda_i$ has a tail that decays slower than an exponential rate and thus allows $w_i$ not to be shrunk. A horseshoe prior assumes half-Cauchy distribution over $\lambda_i$ and $\tau$
\[
\lambda_i \sim C^+(0,1),~~~\tau\sim C^+(0,1).
\]
Being constant at the origin, the half-Cauchy prior has nice risk properties near the origin~\citep{polson2009alternative}. \citet{polson2010shrink} warn against using empirical Bayes or cross-validation approaches to estimate $\tau$, due to the fact that MLE estimate of $\tau$ is always in danger of collapsing to the degenerate $\hat \tau  = 0$~\citep{tiao1965bayesian}.

A feature of the horseshoe prior is that it possesses both tail-robustness and sparse-robustness properties~\citep{bhadra2017horseshoe+}; meaning that  an infinite spike at the origin and very heavy tail  that still ensures integrability. The horseshoe prior can also be specified as
\[
w_i\mid\lambda_i,\tau \sim N(0,\lambda_i^2),~~~\lambda_i\mid \tau \sim C^+(0,\tau),~~~\tau\sim C^+(0,1)
\]

The log-prior of the horseshoe cannot be calculated analytically, but a tight lower bound  ~\citep{carvalho2010horseshoe} can be used instead
\begin{equation}
\phi_{HS} ( w_i | \tau ) = - \log p_{HS} ( w_i | \tau ) \ge - \log \log \left ( 1 + \frac{2 \tau^2}{w_i^2} \right ) .
\end{equation}
The motivation for the horseshoe penalty arises from the analysis of the prior mass and influence on the posterior in {\bf both} the tail and behavior at the origin. The latter provides the key determinate of the sparsity properties of the estimator. 


When Metropolis-Hasting MCMC is applied to horseshoe regression, it suffers from sampling issues. The funnel shape geometry of the horseshoe prior is makes it challenging for MCMC to efficiently explore the parameter space. \citet{piironen2017sparsity} proposed to replace Cauchy prior with half-t prior with small degrees of freedom and showed improved convergence behavior for NUTS sampler~\citet{hoffman2014no}. \citet{makalic2016simple} proposed using a scale mixture representation of half-Cauchy which leads to conjugate hierarchy and allows a Gibbs sample to be used.  \citet{johndrow2017scalable} proposed two MCMC algorithms to calculate posteriors for horseshoe  priors. The first algorithm addresses computational cost problem in high dimensions by approximating matrix-matrix multiplication operations. For further details on computational issues and packages for horseshoe sampling, see~\citet{bhadra2017lasso}. An issue of high dimensionality was also addressed by~\citet{bhattacharya2016fast}.

One approach is to replace the thick-tailed half-Cauchy prior over $\lambda_j$ with half-t priors using small degrees of freedom. This leads to the sparsity-sampling efficiency trade-off problem. Larger degrees of freedom for a half-t distribution will lead to more efficient sampling algorithms, but will be less sparsity inducing. For cases with large degrees of freedom, tails of half-t are slimmer and we are required to choose large $\tau$ to accommodate large signals. However, priors with a large $\tau$ are not able to shrink coefficients towards zero as much.

\subsection{Spike-and-slab Prior}
Spike-and-slab is another sparsity inducing prior widely used in Bayesian analysis. It assumes that the prior is a mixture of point-mass $\delta$ distribution and Gaussian distribution~\cite{polson2017bayesian}
\[
w_i \mid a, \sigma^2 \sim (1-a)\delta_0 + a N\left(0, \sigma^2\right) .	
\]
Here $a\in \left(0, 1\right)$ controls the overall sparsity in $w$ and $\sigma^2$ allows for  non-zero weights. By setting  $w_i  =  \gamma_i\alpha_i$, we get a Bernoulli-Gaussian mixture model given by
\begin{equation}
\label{eq:bg}
\begin{array}{rcl}
\gamma_i\mid a & \sim & \text{Bernoulli}(a) \ ;
\\
\alpha_i\mid \sigma^2 &\sim & N\left(0, \sigma^2\right) \ .
\\
\end{array}
\end{equation}
Since $\gamma$ and $\alpha$ are independent, we can write the joint density function as a product
$$
\begin{array}{rcl}
p\left(\gamma, \alpha \mid a, \sigma^2\right) & = & \prod\limits_{i = 1}^p p\left(\alpha_i, \gamma_i \mid a, \sigma^2\right) \\
& = & 
a^{\|\gamma\|_0}
\left(1-a\right)^{p - \|\gamma\|_0}
\left(2\pi\sigma^2\right)^{-\frac p2}\exp\left\{-\frac1{2\sigma^2}\sum\limits_{i = 1}^p\alpha_i^2\right\} \ .
\end{array}
$$
Here $\lVert\gamma\rVert_0 = \sum_i I(\gamma_i \ne 0)$ is the number of non-zero entries in the vector $\gamma$, and $p$ is the length of vector $w$. It can be shown that finding the MAP estimator for the linear model given by Equation \ref{eq:bayes1} with Spike-and-Slab prior is equivalent to solving the following optimization problem for $\gamma$ and $\alpha$ \cite{soussen2011bernoulli,polson2017bayesian}
\[
\minf_{\gamma, \alpha}\left\|\tilde{\boldsymbol{\rho}} - R_\gamma \alpha_\gamma\right\|_2^2 + \dfrac{\sigma_e^2}{\sigma^2}\left\|\alpha\right\|_2^2 + 2\sigma_e^2\log\left(\frac{1-a}{a}\right) \left\|\gamma\right\|_0.
\]
Here $R_\gamma \defeq \left[R_i\right]_{i \in S}$ is the matrix with columns that have index inside set $S$, and $S$ is the set of ``active explanatory variables"  with $S = \{i: \gamma_i = 1\}$ and $\alpha_\gamma \defeq \left(\alpha_i\right)_{i \in S}$ be their corresponding coefficients. 

\subsection{Laplace Prior (Lasso)}
A double exponential (Laplace) prior distribution \citep{carlin1991inference} for each weight $w_i$ was previously shown to be an effective mechanism to regularize the portfolio \citep{brodie2009sparse} and to incorporate investor's preferences for the number of assets in the optimal portfolio \citep{puelz2015sparse}.
\[
p(w_i \mid b) = 1/(2b)\exp(-|w_i|/b).
\]
The log-posterior is then given by
\[
\log p(w \mid R,\tilde{\rho}, b) \propto \lVert \tilde{\rho}-Rw\rVert_2^2 + \dfrac{\sigma^2_e}{b}\lVert w\rVert_1.
\]
For $b>0$, the posterior mode is equivalent to the $\ell_1$-penalized estimate with $\lambda = \sigma^2_e/b$. Large variance $b$ of the prior is equivalent to the small penalty weight $\lambda$ in the $\ell_1$-penalized objective function.


\cite{park2008bayesian} represent the Laplace prior is a scale Normal mixture to develop a Gibbs sampler that iteratively samples  $w$ and $b$ to estimate joint distribution over $(\hat w, \hat b)$. Thus, we do not need to apply cross-validation to find optimal value of $b$, the Bayesian algorithm calculates it  automatically. 

Given data $D = (R,\tilde{\rho})$, where $R$ is the  $n\times p$ matrix of standardized regressors and $y$ is the $n$-vector of outputs.  Implement a Gibbs sampler for this model when Laplace prior is used for model coefficients $w_i$. Use scale mixture normal representation.
\begin{align*}
w \mid  \sigma^2,\tau_1,\ldots,\tau_p \sim  & N(0,\sigma^2D_{\tau})\\
D_{\tau} = & \mathrm{diag}(\tau_1^2,\ldots,\tau_p^2)\\
\tau_i^2  \mid \lambda \sim &Exp (\lambda^2/2)\\
p(\sigma^2) \propto & 1/\sigma^2.
\end{align*}
Then the complete conditional required for Gibbs sampling are given by
\begin{align*}
w \mid D,D_{\tau} \sim & N(A^{-1}R^Ty,\sigma^2A^{-1}), ~~A = R^TR+ D^{-1}_{\tau}\\
\sigma^2 \mid w,D,D_{\tau} \sim & \mathrm{InverseGamma}\left((n-1)/2+p/2,(\tilde{\rho}-Rw)^T(\tilde{\rho}-Rw)/2 + w^TD_{\tau}^{-1}w/2\right)\\
1/\tau_j^2 \mid w_j,\lambda \sim & \mathrm{InverseGaussian}\left(\sqrt{\dfrac{\lambda^2\sigma^2}{w_j^2}},\lambda^2\right)
\end{align*}
The formulas above assume that  $R$ is standardized, e.g. observations for each feature are scaled to be of mean zero and standard deviation one, and $\tilde{\rho}$ is assumed to be centered. 

We can use initialize the parameters as follows
\begin{align*}
w = & (R^TR + I)^{-1}R^T\tilde{\rho}\\
q = & \tilde{\rho} - Rw\\
\sigma^2 = & q^Tq/n\\
\tau^{-2} = &1/w^2\\
\lambda  = &  p  \sqrt{\sigma^2} / \lVert w \rVert_1.
\end{align*}
where $n$ is number of  observations and $p$ is number of columns (inputs) in matrix $X$.

There are several efficient optimization algorithms to compute mode of the posterior distribution for a Laplace prior. The most widely used approaches in applications  are LARS \citep{efron2004least} and coordinate descent \citep{friedman2010regularization}. The advantage of LARS compared to other optimization techniques is that it provides a way to compute the sequence of solutions for different values of the penalty weight $2\sigma^{2}/b$. Coordinate descent algorithm which updates one parameter at a time, holding the others fixed was shown to be more computationally efficient. 

Finally, it is important to note the posterior contraction properties of all of these sparsity priors. Not all of the priors have similar minimax rates and properties.
For example, lasso can underperform priors such as the horseshoe. For example, 
\citep{castillo2015bayesian} analyses Bayesian linear regression with sparsity-inducing prior of the following form
\[
(S,\beta) \sim \pi_p(|S|)\left(\begin{array}{c}
p\\|S|	
\end{array}\right)^{-1}g_s(\beta_S)\delta_0(\beta_{\bar S})
\]
here $S \subset \{1,\dots,p\}$, is a random subset of cardinality $|S|$ and $\beta_S = \{\beta_i\;:\; i\in S\}$, and $\bar S = \{1,\dots,p\} - S$. The prior $\pi_p$ performs the model selection by inducing sparsity and $g_s$ density model only non-zero coordinates. \cite{song2017nearly} extend this analysis and to the cases when shrinkage priors have heavy and fat tail and also show how estimation algorithm for hierarchical models with spars-inducing priors adopt to  unknown $\sigma_e$.

Spike and slab priors are a special case of this prior with $\pi_p$ being the Binomial and $g_s$ being the product of Laplace of Normal univariate distributions. 

They show that under certain conditions on the design matrix, the posterior distribution is shown to contract at the optimal rate for recovery of the unknown sparse vector, and to give optimal prediction of the response vector, when $g_S$ is the product of Laplace densities $(1/2)\lambda\exp(-\lambda|\beta|)$ and the inverse scale parameter is bounded by $\lVert X \rVert/p \le \lambda \le 2\bar \lambda$, where $\bar \lambda = 2 \lVert X \rVert\sqrt{\log p}$, with $\lVert X \rVert^2 = \max_i \text{trace}(R^TR)_i$.

\cite{van2014horseshoe} show that the $\tau$ parameter of the horseshoe can be interpreted as the proportion of nonzero weights up to a logarithmic factor. Further, they show that when true sparsity level $p_n$ is know and $\tau$ is set to $(p_n/n)\sqrt{\log(n/p_n)}$ then the horseshoe estimator attains the minimax $\ell_2$ risk and posterior contraction is bounded above by this number, when $p_n$ is estimated from data (empirical Bayes) or a hierarchical prior is used \citep{van2017adaptive}. 

\cite{van2017uncertainty} show that the horseshoe estimator (the posterior mean) leads to credible balls and marginal credible intervals that have an optimal size if the sparsity level of the prior is set correctly. Their proofs are derived for the posterior under the sparse normal means problem.

\section{Applications}
Our sparse portfolio selection model is to construct optimal portfolios made by three models with sparsity-inducing priors; Laplace, horseshoe \citep{carvalho2010horseshoe}, and spike-and-slab. The LARS algorithm \cite{efron2004least} is used to find posterior mode for the Laplace model, MCMC to generate samples form the horseshoe \cite{hahn2019efficient} model, and Single Best Replacement (SBR) \cite{polson2017bayesian} algorithm to find the posterior mode of the spike-and-slab.

We demonstrate how Laplace regularized portfolios allocations and the corresponding LARS algorithms leads to an intuitive way to select an optimal portfolio and assign a selection order of the stocks to be included into the portfolio. We use daily returns from three different portfolios. One portfolio corresponds to a widely used stock index (SP100) and two portfolios of stocks managed by two different hedge funds, namely Viking Global Investors and Renaissance Technologies. We use the top 50 holdings of each portfolio and apply our selection algorithm to design a sparse portfolio with minimal risk level while guaranteeing to out-perform the SP500 index. We use daily returns during the period from 2016-02-23 to 2018-02-15 (500 trading days) as our training data and returns for the period 2018-02-16 to 2019-02-22 for calculating out-of-sample performance of our portfolios. An optimal penalty  parameter $\lambda$ can be calculated using cross-validation, but we also show that output of the LARS algorithm allows us to infer  $\lambda $ given the desired number of assets in the portfolio. 

\subsection{Small-sized Stock Portfolio}
First, we demonstrate how the LARS algorithm finds the posterior mode  under $\ell_1$ regularization and also provides a natural ranking of the importance of individual stocks. At every step of the LARS algorithm, a new variable enters the active set and thus it performs the same number of steps as the number of variables. The order in which LARS adds the variables corresponds to their importance, implying that the variables added in the beginning lead to a model that fits the training data well and of a low variance. 

\begin{figure}[H]
	\begin{tabular}{p{1em}m{0.5\linewidth}m{0.5\linewidth}}
	 \rotatebox{90}{Viking} & \includegraphics[width=\linewidth]{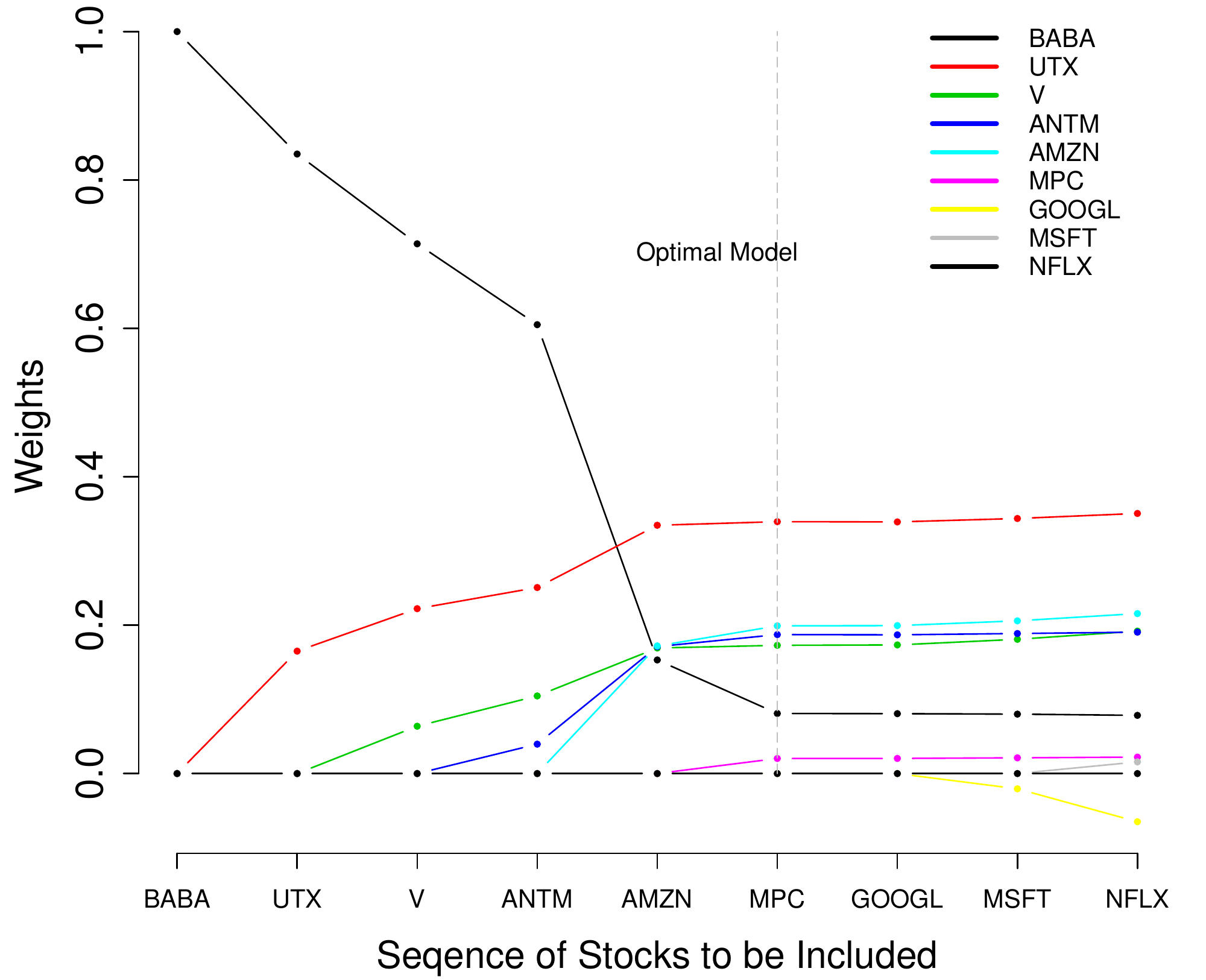} & \includegraphics[width=\linewidth]{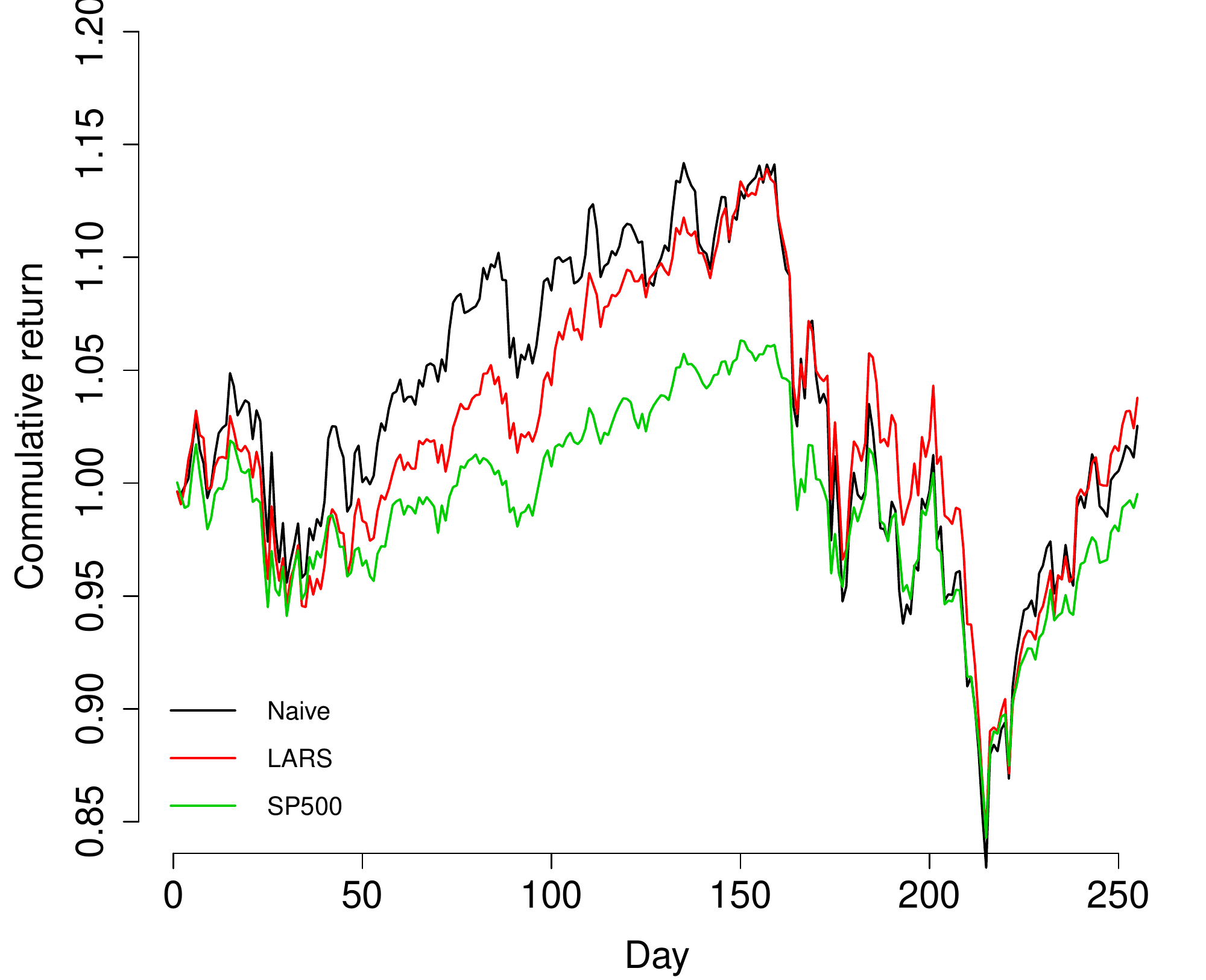}\\
	\rotatebox{90}{Renaissance}&\includegraphics[width=\linewidth]{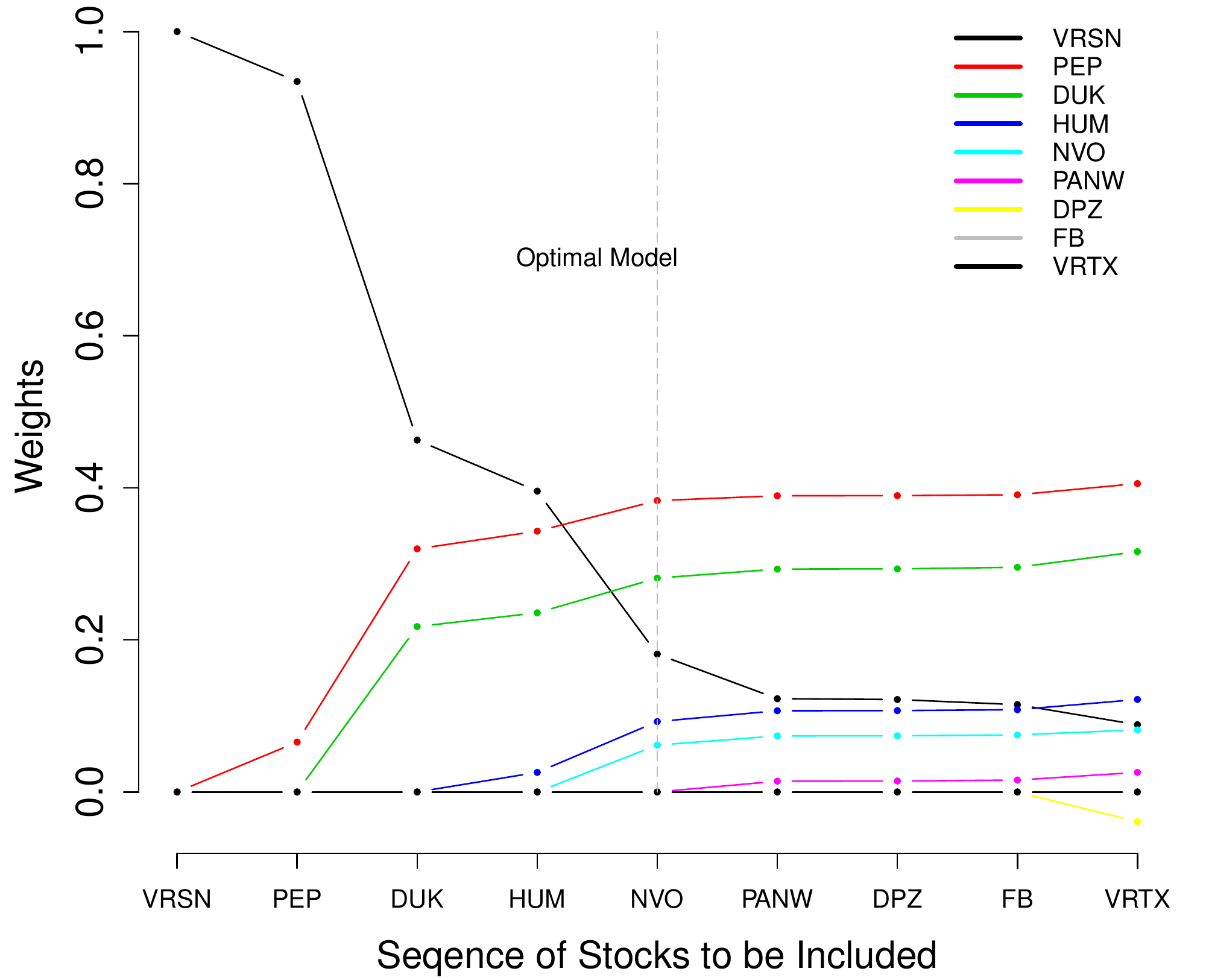} & \includegraphics[width=\linewidth]{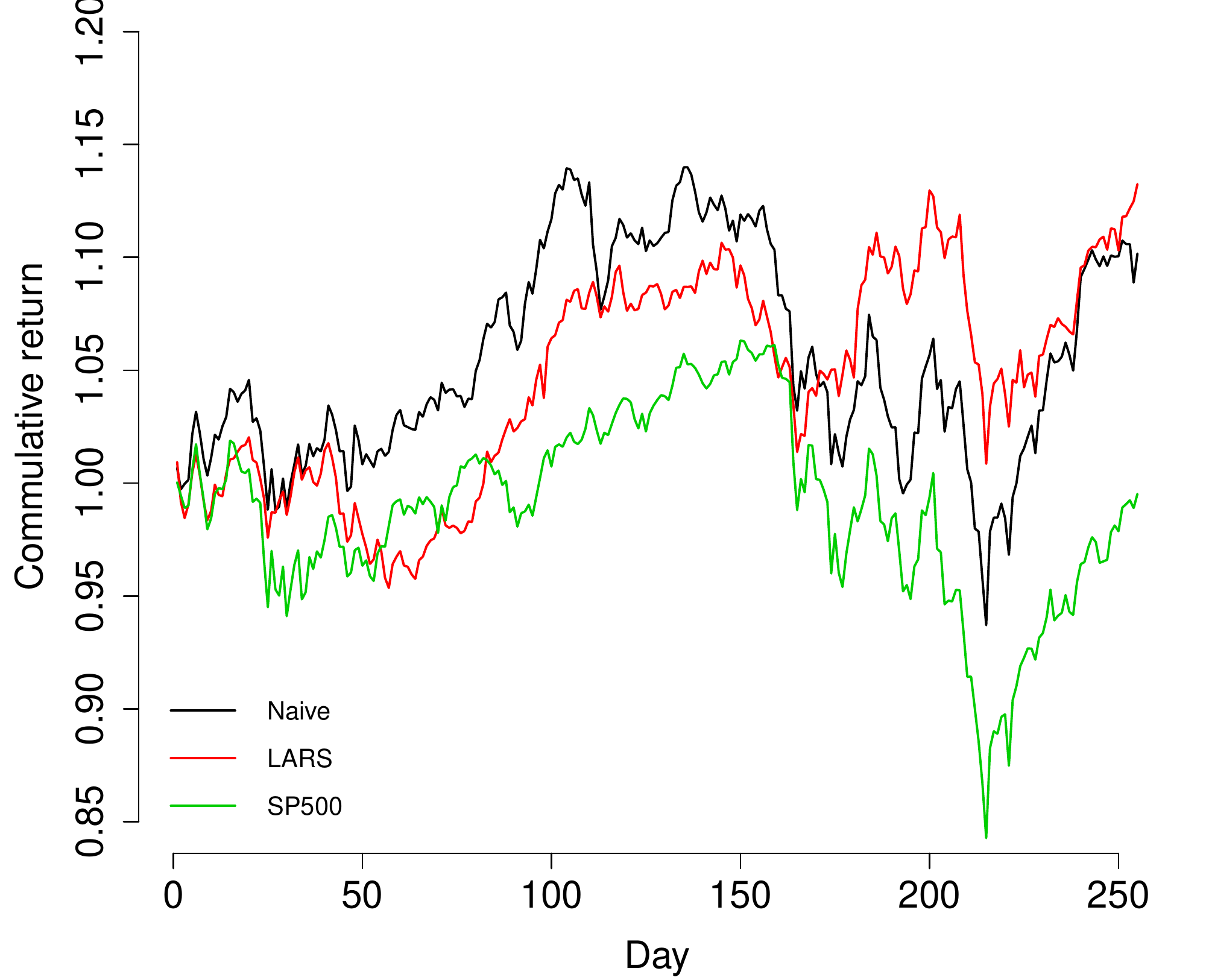} \\
	\rotatebox{90}{SP100}&\includegraphics[width=\linewidth]{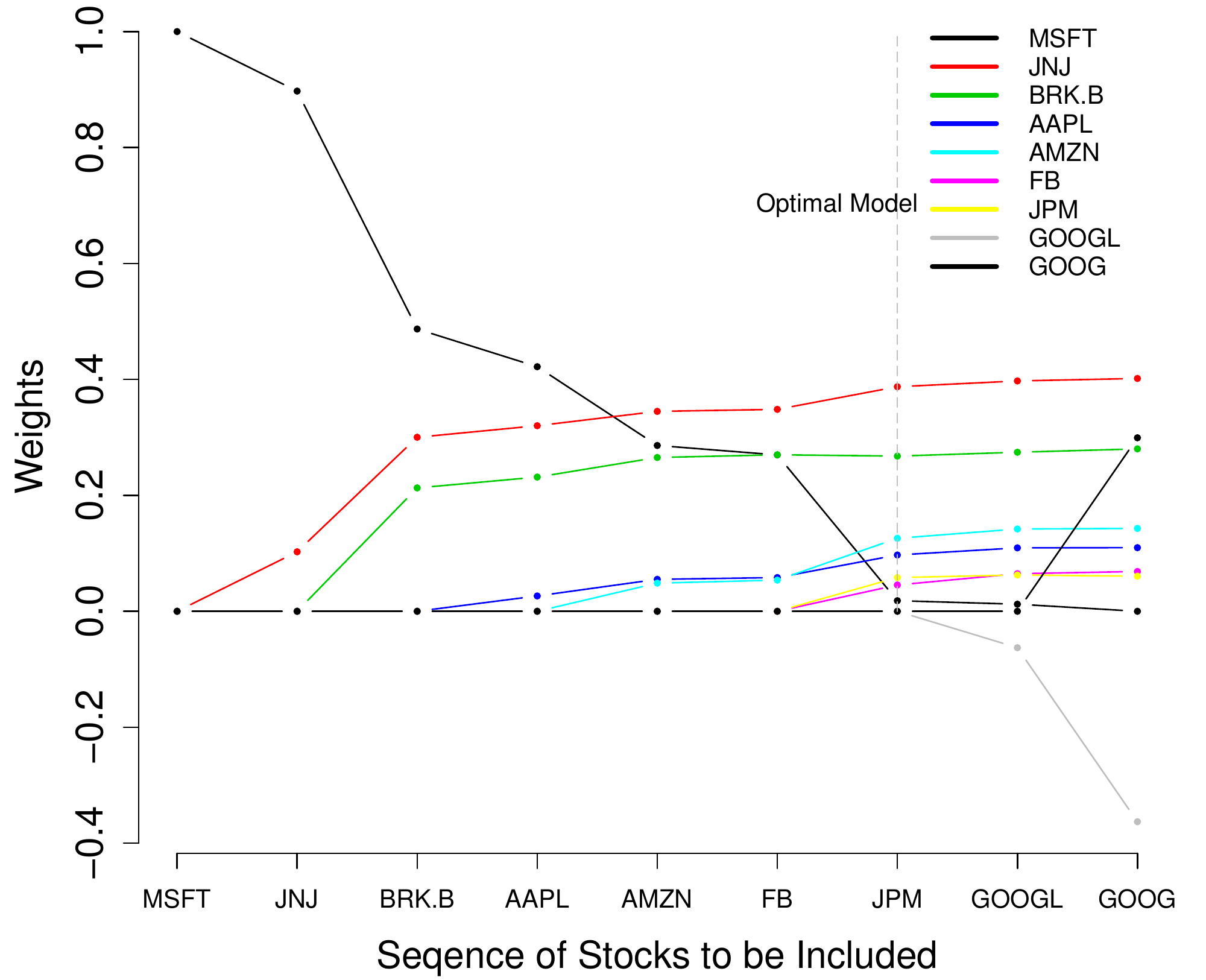} & \includegraphics[width=\linewidth]{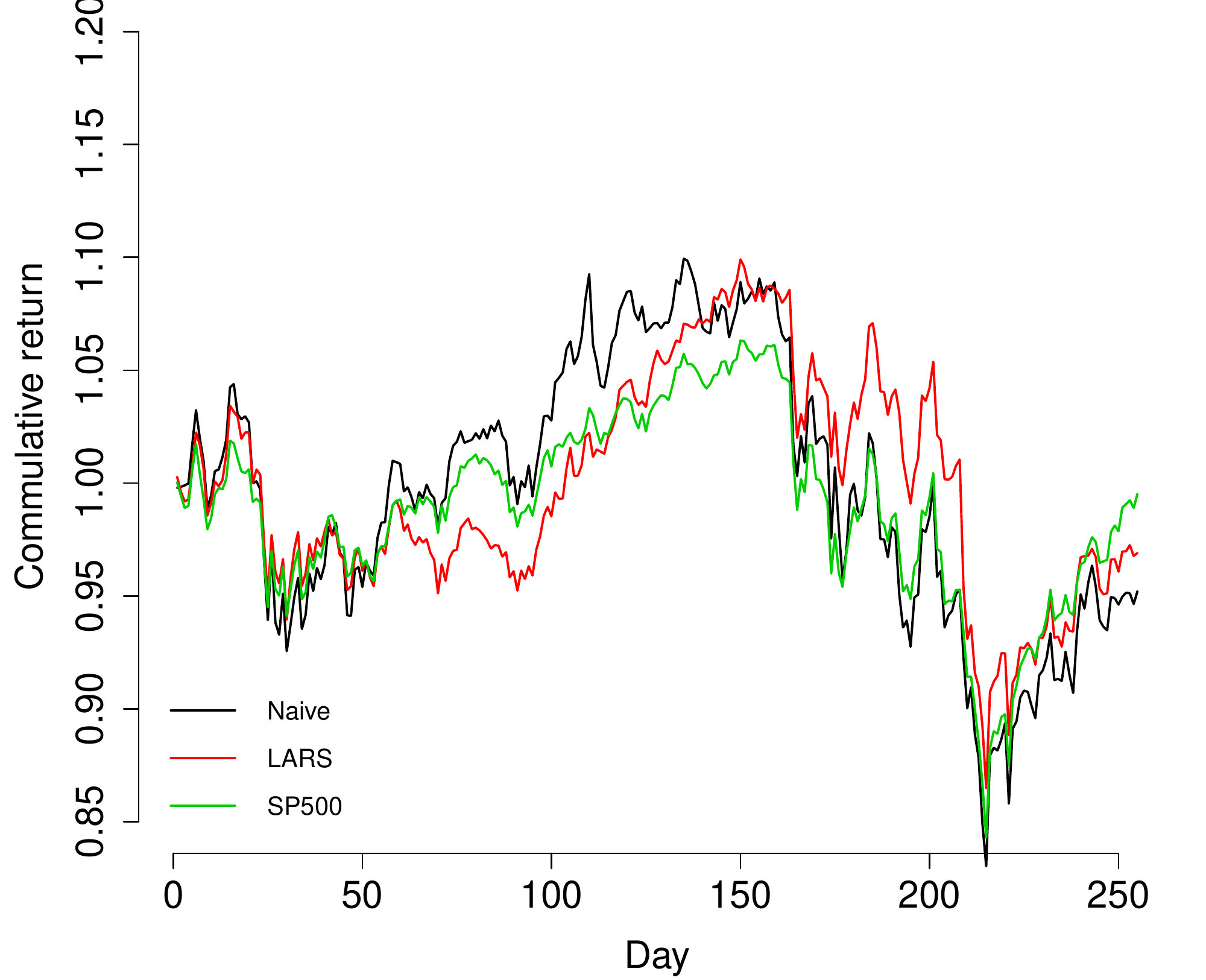}\\
	&\;\;\;\;(a) Sequence of stocks & \;\;\;\;(b) Out-of-sample cumulative return
\end{tabular}
	\caption{(a) Sequence of stocks to be added to the portfolio as the function of number of stocks allowed (b) Out-of-sample cumulative return of an optimal LARS portfolio compared with naive portfolio and SP500}
	\label{fig:lasso-selection}
\end{figure}

To empirically evaluate our selection procedure we take the top nine holdings from the SP100, Renaissance and Viking Global. Then the LARS algorithm  adds one stock at a time to the portfolio and evaluates the out-of-sample performance at each time. We select an optimal portfolio that leads to the best out-of-sample returns. Figure~\ref{fig:lasso-selection}(a) shows the weights assigned by the LARS algorithm at each iteration and the step at which the optimal portfolio was achieved. 

Visualizing outputs of the LARS algorithms gives an inventor a way to interpret the importance of each of the stocks in the portfolio and suggest modifications. If an investor wants a smaller portfolio he/she can remove stocks added later by the algorithm.

The next question is whether the LARS selected portfolio does out-perform the naive equally weighted portfolio or traditional Markowitz portfolio selected by solving problem \ref{eqn:markowitz}. We compare our optimal portfolio to performance of the SP500 index. Figure \ref{fig:lasso-selection}(b) shows the cumulative return (growth of \$1 invested) of the LARS selected optimal portfolio and compares it with the naive and SP500 portfolios. 

Table \ref{tab:lasso-small} shows the out-of-sample mean and standard deviation of the returns as well as Sharpe ratio for the  selected portfolios.

\begin{table}[H] \centering 
  \begin{tabular}{@{\extracolsep{-3pt}} l|ccc|ccc|ccc|c} 
 & \multicolumn{3}{c}{Viking} & \multicolumn{3}{c}{Renaissance} & \multicolumn{3}{c}{SP100} &\\
& Naive &	LARS &	QP& Naive &	LARS &	QP& Naive &	LARS &	QP&SP500\\\hline
 $\mu$        & 0.066 & 0.12 & 0.12 & 0.1 & 0.099  & 0.065 & -0.029 & -0.011 & -0.038 & 0.0086 \\
 $\sigma$          & 4.033 & 3.67 & 3.65 & 2.8 & 2.383  & 2.349 & 3.779  & 3.467  & 3.476  & 2.6004 \\
 $\mu/\sigma$ & 4.124 & 7.94 & 8.37 & 9.3 & 10.466 & 7.026 & -1.95  & -0.764 & -2.787 & 0.8376
  \end{tabular} 
  \caption{Out of sample performance of LARS selected portfolio compared with Naive portfolio and SP500} 
  \label{tab:lasso-small} 
  \end{table}

		

The LARS algorithms selected 5 stocks (VRSN, PEP, DUK, HUM and NVO) and the resulting portfolio is less risky and has the same average return when compared to the naive portfolio. 

Plots shown in Figure \ref{fig:lasso-selection} allow us to visualize the  ranking of the stocks in the portfolio and to allow investor to decide how to increase or decrease the number of positions in a portfolio. Further,  the optimal allocations calculated by  LARS lead to a portfolio with lower risk (standard deviation of 0.008) and of higher return (mean of 0.0005) when compared to SP500 and naive  equal weights  allocations. 



\subsection{Large-sized Stock Portfolio}
To demonstrate further, how our portfolio allocation algorithm can be used for selecting from a larger sets of stocks, we again compare models with Laplace ($\ell_1$), Horseshoe and $\ell_0$ regularization. We compare the shrinkage effect and empirical out-of-sample performance of these three selection approaches.

The question is whether the weight shrinkage introduced by the LARS algorithm does effect the portfolio performance and whether Horseshoe or $\ell_0$ selectors lead to sparser portfolios. We select portfolios using the LARS for Laplace prior ($\ell_1$), MCMC algorithm for Horseshoe prior, and Single Best Replacement (SBR) algorithm for Spike-and-Slab prior ($\ell_0$), and non-regularized least-squares approach. We apply all four algorithms to select portfolio from top 35 holdings of Viking and Renaissance hedge funds as well as from SP100 stocks. 

\begin{figure}[H]
	\centering
	\begin{tabular}{p{1em}m{0.82\linewidth}}
	\rotatebox{90}{Viking} &\includegraphics[width=\linewidth]{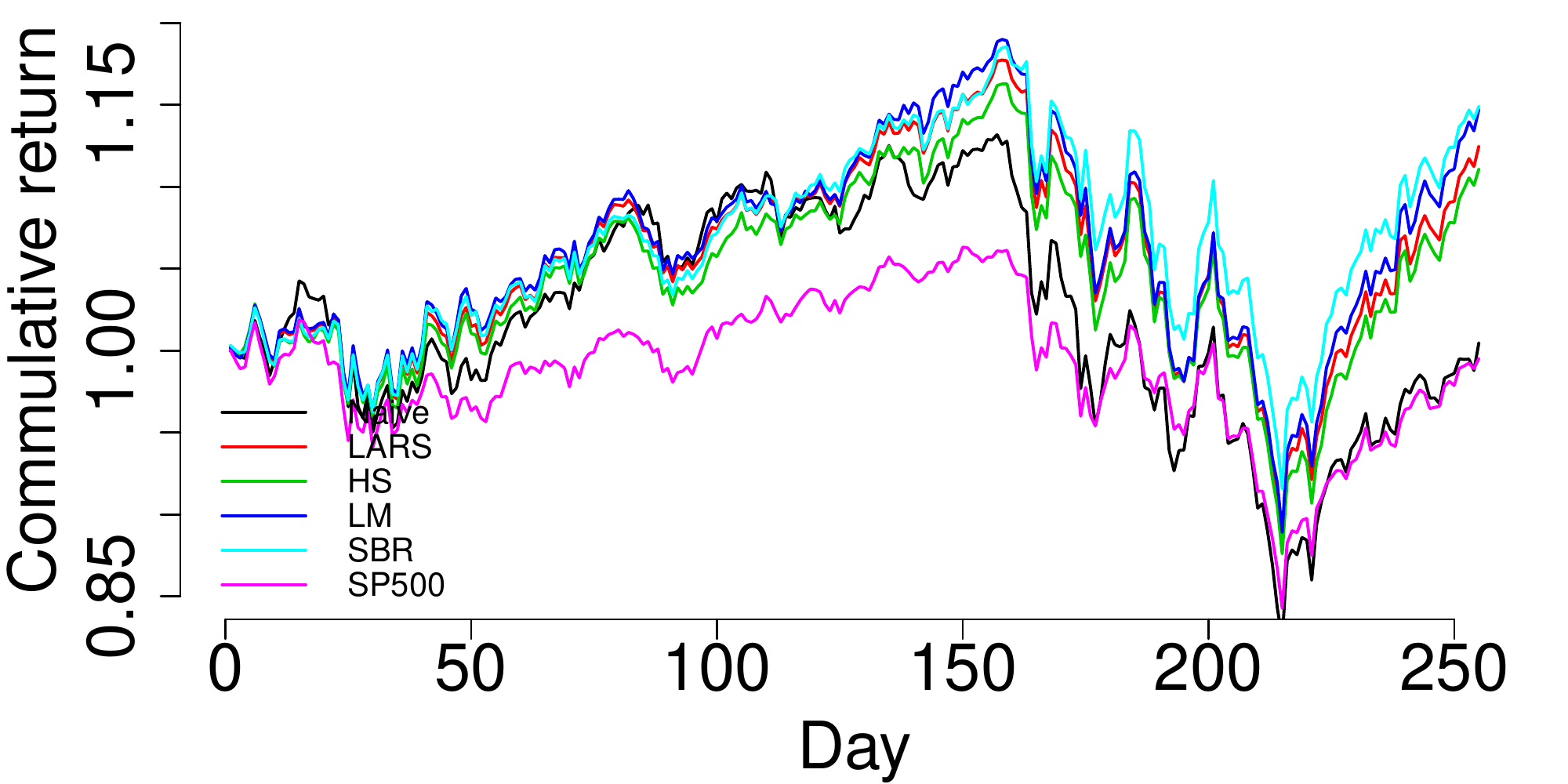} \\
	\rotatebox{90}{Renaissance} &\includegraphics[width=\linewidth]{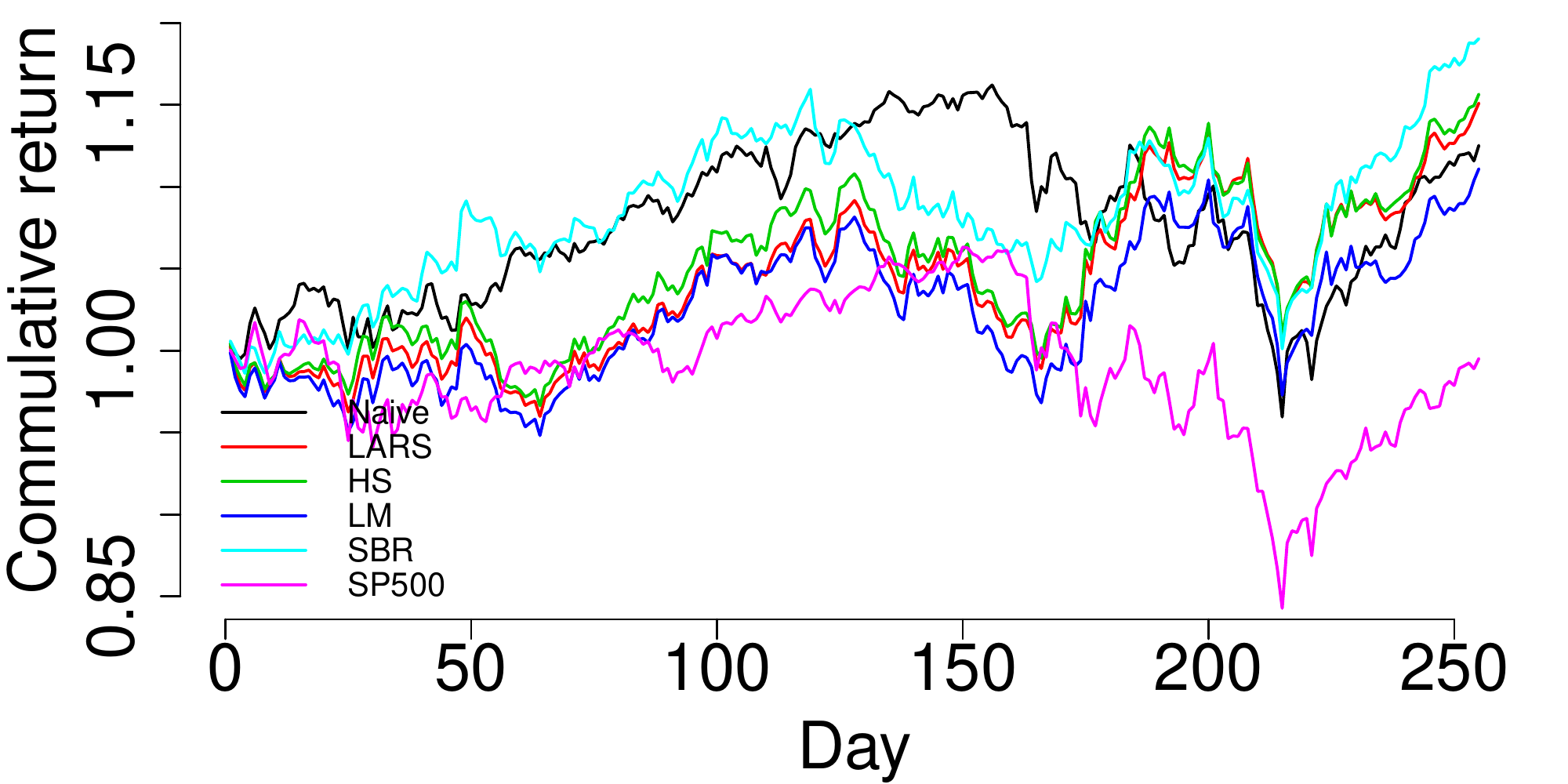}  \\
	\rotatebox{90}{SP100} & \includegraphics[width=\linewidth]{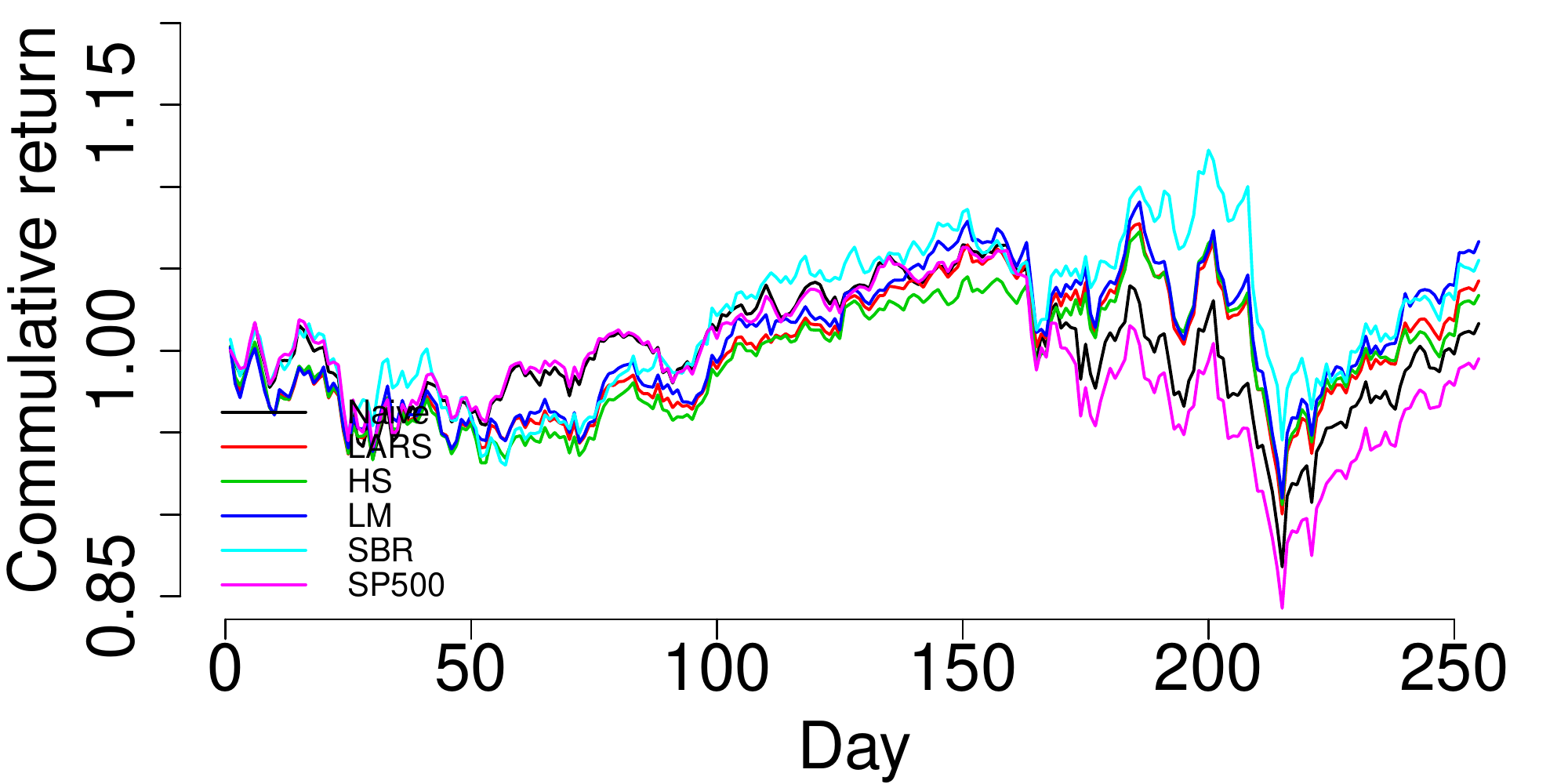}\\
	
	\end{tabular}
	\caption{Out-of-sample cumulative return of an optimal LARS portfolio compared with naive portfolio and SP500}
	\label{fig:all-ret}
\end{figure}

Table \ref{tab:all-perfm} shows the out-of-sample mean ($\mu$), standard deviation ($\sigma$), and sharpe ratio ($\mu/\sigma$) of the daily returns multiplied by 252 as well as the number of stocks selected ($\lVert w \rVert_0$). 
Figure \ref{fig:all-ret} shows the cumulative return.
\begin{table}[!htbp] \centering 
  \begin{tabular}{@{\extracolsep{-3pt}}l|llll|llll|llll} 
 & \multicolumn{4}{c}{Viking} & \multicolumn{4}{c}{Renaissance} & \multicolumn{4}{c}{SP100}\\
 & LARS&	HS&	LM&	L0		  &LARS&	HS&	LM&	L0				&	LARS&	HS&	LM&	L0\\ \hline
 $\mu$        &  0.13  & 0.14  & 0.16  & 0.16  & 0.14  & 0.17 & 0.11  & 0.18  & 0.051 & 0.055 & 0.077 & 0.066 \\
 $\sigma$      &3.21  & 3.2   & 3.3   & 3.23  & 2.19  & 2.14 & 2.26  & 2.05  & 2.401 & 2.386 & 2.533 & 2.52  \\
 $\mu/\sigma$  & 10.45 & 10.96 & 12.02 & 12.34 & 16.47 & 20.4 & 12.72 & 22.15 & 5.39  & 5.779 & 7.613 & 6.578 \\
 $\lVert w \rVert_0$ &22    & 11    & 31    & 11    & 25    & 12   & 32    & 4     & 16    & 9     & 32    & 4   
  \end{tabular} 
  \caption{Out of sample performance of LARS selected portfolio compared with Naive portfolio and SP500} 
  \label{tab:all-perfm} 
  \end{table} 

Horseshoe  and $\ell_0$ selectors do out-perform the LARS selector and lead to sparser portfolios. From practical standpoint, the $\ell_0$ selector not only leads to the best performing and the most sparse portfolio, it is also the easier to use when compared to the horseshoe selector. $\ell_0$ requires investor to specify one parameter instead of two as in the horseshoe and the penalty term in $\ell_0$ is arguably more interpretable. 


\section{Discussion}
The goal of our paper is to  present an extension of the Black-Litterman portfolio framework. A quadratic programming optimization problem is recast as an inference problem for hierarchical Bayesian linear model. We have shown how linear constraints of the optimization problem can be formulated as the product of exponential priors and sparsity priors. The main advantage of our formulation is the ability to incorporate an investor's subjective opinion about which assets to be included into the portfolio. Specifically, we demonstrated how sparsity priors stabilize the portfolio selection and to select a small number of stocks to be included into the portfolio. The sparsity priors  correspond to investors' preferences for portfolios with small number of stocks. We used our hierarchical formulation to demonstrate empirical performance of several sparsity-inducing priors versus the SP100 index. We have shown that Horseshoe and Spike-and-Slab ($\ell_0$ penalty)  priors not only lead to portfolios with smaller number of stocks but also have better out-of-sample performance when compared to Laplace prior ($\ell_1$ penalty) and traditional Markowitz (un-regularized) portfolio selection procedure.  Inclusion of short positions in regularized portfolio also leads to better performance and lower risk while maintaining stability of the portfolio (no extremely large weights). From a practical standpoint $\ell_0$ penalty leads to the the best performing portfolio and requires an investor to specify only one parameter. 

There are a number of direction for future research. For example, equivalence between non-convex penalties and hierarchical linear models, as well as further study of elastic net formulations of the Black-Litterman framework.
\bibliography{ref}
\end{document}